\documentclass[10pt, conference, compsocconf]{IEEEtran}
\usepackage{epsfig}
\usepackage{url}
\usepackage{listings}
\usepackage{dblfloatfix}
\begin{document}
\title{Using Positional Heel-marker Data to More Accurately Calculate Stride Length for Treadmill Walking: A Step Length Approach}

\author{
\IEEEauthorblockN{Kevin Supakkul}
\IEEEauthorblockA{Department of Mechanical Engineering\\
Stanford University\\
Stanford, California, USA\\
	supakkul@stanford.edu}
%\and
%\IEEEauthorblockN{Lawrence Chung}
%\IEEEauthorblockA{Dept. of Computer Science\\
%Univ. of Texas at Dallas, USA\\
%chung@utdallas.edu}
}

\maketitle

\begin{abstract}

Treadmill walking is a convenient tool for studying the human gait; however, a common gait parameter, stride length, can be difficult to calculate directly because relevant reference points continually move backwards.  Although there is no direct calculation of stride length itself, we can use positional heel-marker data to directly determine a similar parameter, step length, and we can sum two step lengths to result in one stride length.  This proposed method of calculation is simple but seems to be unexplored in other literature, so this paper displays the details of the calculation. Our experimental results differed from the expected values by 2.2\% and had a very low standard deviation, suggesting that this method is viable for practical use.  The ability to calculate stride length for treadmill walking using heel-marker data may allow for quick and accurate gait calculations that further contribute to the versatility of heel data as a tool for gait analysis.

\end{abstract}


\IEEEpeerreviewmaketitle

\section{Introduction}
%context

Treadmills are useful for gait analysis because they are continuous and controllable. Treadmill walking has many benefits, including but not limited to longer trial durations, consistent walking speeds and inclines, and more convenient environments to place sensors and cameras; such a medium for gait analysis has shown to be useful for multiple fields such as rehabilitation \cite{abel2002gait} and robotic orthoses \cite{colombo2000treadmill}.

%problem

Since the treadmill is a tool for gait analysis, we must be able to obtain accurate and consistent gait parameters from it.  One useful tool for finding gait parameters, especially spatial parameters, is a positional heel marker manufactured by Vicon which measures the X,Y, and Z coordinates of the heels.  However, even with positional heel markers, spatial parameters - namely stride length - can be difficult to calculate in an accurate manner due to the mechanics of treadmill walking.  Stride length is defined as the distance from heel-strike to heel-strike of one foot \cite{Whittle}.  This is difficult to directly track on a treadmill: if we subtract the location of the heel at the previous heel strike from the location of the current heel strike, the result is close to zero because the foot returns to its original position after each stride (see Figure \ref{Heelmarkergraph} for a graphical representation of heel marker data).  Thus, at the moment of a heel strike, there is no data point we can reference as the positional location of the previous heel strike relative to the current heel strike, making a direct calculation of stride length difficult to obtain.  The closest indication we have of the previous heel strike location is the foot's path of travel during the stride, which is recorded in the Vicon data (Figure \ref{Heelmarkergraph}).  However, although it may seem intuitively sound, taking a total-distance-traveled measurement of the heel does not result in an accurate stride length, as discussed later in the paper.  Therefore this issue of moving reference points is the main challenge in the calculation of stride length from treadmill walking.

Other researchers have calculated stride length values from treadmill walking.  In general, many prior methods that we found are usable but contain key concerns that can be improved.
One previously used method involves video analysis of the treadmill walk \cite{Murray87} \cite{Wall81} \cite{padulo2014}.  Murray et al. analyzed video recordings by tracing reference points on paper to determine spatial gait parameters \cite{Murray87}.  This method is very sound and is perhaps the most accurate way to measure spatial parameters for treadmill walking; however, one drawback is that this method requires manual calculations of each stride. 
This manual video-based method is slower and less convenient than digital algorithm-based methods, especially if a researcher's desired analysis requires stride length calculations from tens or hundreds of strides.

Padulo et al., in addition to mentioning video analysis, also provided an alternative method for stride length calculation \cite{padulo2014}.  After calibrating treadmill speed, one can obtain stride length after finding stride frequency.  However, this requires an additional manual step in calibrating treadmill speed, and there may be differences in walking speed from stride to stride \cite{Souza16} that cannot be determined from a pre-walk calibration, thus affecting stride length results.  Qi et al. also employed a digital stride length calculation \cite{qi2014estimation}; however, their main focus was on the accuracy of sensor hardware and not the stride length algorithm, so they did not test their stride length calculations with a control specific to stride length.  Their algorithm for stride length estimation is similar to one we discuss in Section \ref{sec-discussions}, which we demonstrate is not the most accurate.

Some sources that mentioned stride length in regards to treadmill walking did not explicitly state their algorithm for calculating this parameter \cite{alton1998kinematic} \cite{riley2006kinematic}.  Riley et al. also used Vicon Plug-in Gait data to calculate stride length.  Although their stride length values looked consistent with the other parameters, they mentioned that their data was pre-processed and thus did not provide specific stride length algorithms in the paper \cite{riley2006kinematic}.  Alton et al. used ankle and toe position markers \cite{alton1998kinematic}, similar to Vicon heel markers, but they did not share a detailed algorithm and their results had a slight inconsistency.  They studied the relationship between grounded walking and treadmill walking by first having subjects walk on the ground then on a treadmill set to the average speed of their grounded walking (to the nearest .2 m/s).  Therefore the walking speed calculated from the treadmill data should be close or equal to the walking speed obtained from grounded walking.  We calculated walking speed from their all-subjects data table using their stride lengths and swing/stance times, and we found that the average treadmill walking speed was 1.40 m/s while the average grounded walking speed was 1.28 m/s (a difference of over .4 km/hr), meaning that their treadmill walking speed has a not-insignificant 9.1 percent error. Based on our reverse engineering from their reported data and descriptions, we speculate it is possible that the stride length calculation they used is one that is discussed later in this paper (Section \ref{sec-discussions}): calculating stride length as the total distance traveled by the heel/foot during one stride.  As later discussed in Section \ref{sec-discussions}, although this method seems the most definition-based and the most intuitive, it is faulty for treadmill walking, resulting in stride lengths and walking speeds that are too large due to how double stance time affects positional data during treadmill walking.

This paper addresses some of the aforementioned concerns; our proposed method is digital so it can be performed on large or even real-time sets of data, our results show a 2.2\% error which is small enough to suggest that this method is practical, and this paper contains specific details about the stride length calculation.  Our method of calculation is to sum two consecutive step lengths, with such step lengths to be calculated from horizontal heel-marker data. Since both feet are planted during a heel-strike moment, there exist reference points to perform a direct step length calculation, and the sum of two consecutive step lengths is geometrically equal to one stride length.  This proposed calculation is uncomplicated and rather basic, but we did not see it in other literature and thus decided to explore the theory and experimental numbers behind this method as well as highlight a possible misinterpretation of heel-marker data.

This paper is organized as follows: Section \ref{sec-solution} presents our approach to determining stride length for treadmill walking, Section \ref{sec-experiment} elaborates on our experimental methods and data, Section \ref{sec-discussions} discusses limitations of the study and investigates a possible misinterpretation of heel-marker data, and Section \ref{sec-conclusion} concludes the paper.

\section{Solution: A Step Length Approach}
\label{sec-solution}

\begin{figure}[h]
	\centering
    \epsfig{file=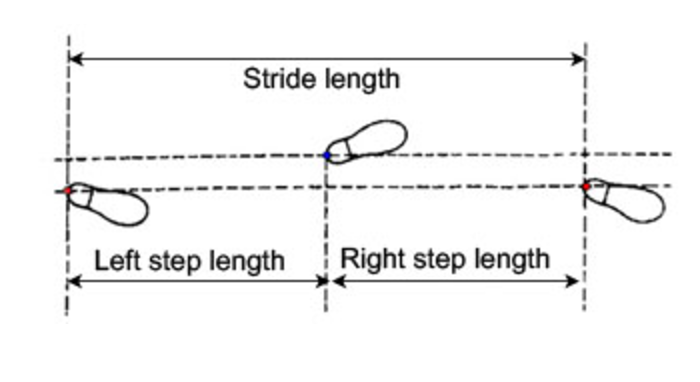,width=3in}
    \caption{Image from \cite{stepdiagram} illustrating the geometry of a stride length.  Two consecutive step lengths are equivalent to one stride length.}
    \label{stepdiagram}
\end{figure}

\begin{figure}
	\centering
	\epsfig{file=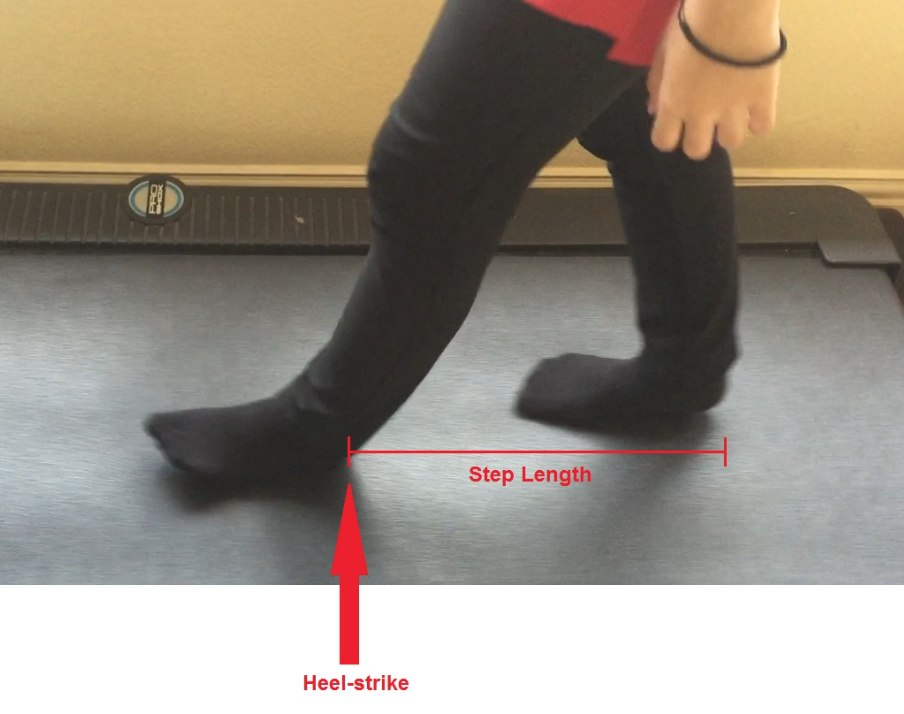, width=2.8in}
	\caption{At the moment of heel strike the two feet are on the ground and are stationary relative to each other.  We can subtract the horizontal heel-marker distances to get a close measurement of step length.  In this image, the calculated value will be the left foot step length.  We can add this to the next right foot step length to get the stride length for this particular stride.}
	\label{treadmillstep}
\end{figure}

Our proposed method of stride length calculation for treadmill walking is to sum two step lengths. Since both feet are on the ground at the same time during a heel strike, the feet are stationary relative to one other, so we can have a direct calculation of step length by subtracting the location of one heel from the other (see Figure \ref{treadmillstep} for visual representation).  Summing two step lengths to result in a stride length might be seen as an unconventional way to calculate this parameter, since the accepted definition of stride length is heel-to-heel distance of the same foot.  However, the sum of two adjacent step lengths is geometrically identical to a stride length \cite{Whittle}\cite{stepdiagram} (see Figure \ref{stepdiagram}), so the calculated values should be equivalent.

\begin{figure}[h]
	\centering
	\epsfig{file=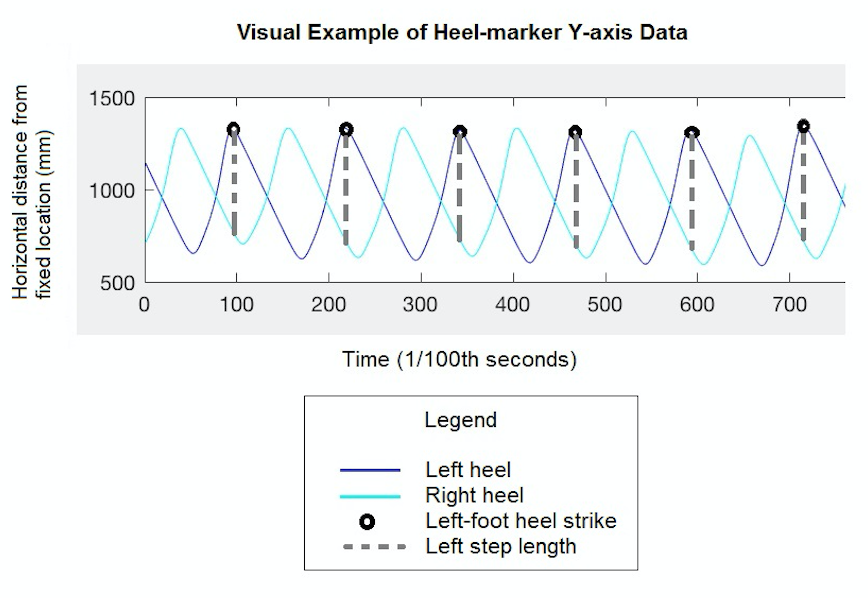, width=3.25in}
	\caption{This plot displays a visual representation of horizontal heel-marker data for treadmill walking.  The data for both feet are graphed together in different colors.  The local maxima represent heel-strikes \cite{BANKS2015101}\cite{DESAILLY200976}, and at these heel-strike moments we calculate step length as the horizontal distance between the two heels.  See Figure \ref{treadmillstep} for photo representation.}
	\label{Heelmarkergraph}
\end{figure}

For our calculations, we focused on the horizontal (y-axis) heel-marker data, which graphically appears as displayed in Figure \ref{Heelmarkergraph}.  As aforementioned, we calculate step length at the point of heel strike, and for filtered horizontal heel-marker data the heel strike moments can be found at the local maxima \cite{BANKS2015101} \cite{DESAILLY200976}.  Thus we calculate step length as the distance between the right and left heels at the moment of a local maximum (see Figure \ref{Heelmarkergraph}).  We then add this step length to the adjacent step length from the opposite foot to calculate the stride length:

$$Stride\ Length = RSL + LSL$$
$$RSL = RH(tRHS) - LH(tRHS)$$
\begin{equation}
LSL = LH(tLHS) - RH(tLHS)
\end{equation}

\noindent
where RSL is right step length, LSL is left step length, tRHS is the time of right-foot heel strike, tLHS is the time of left-foot heel strike, RH(t) is the horizontal location of the right heel at time t, and LH(t) is the horizontal location of the left heel at time t.  Therefore, RH(tRHS) is the horizontal location of the right heel at the time of right-foot heel strike, and so forth.

\section{Experiment}
\label{sec-experiment}

\begin{figure*}
	\centering
	\epsfig{file=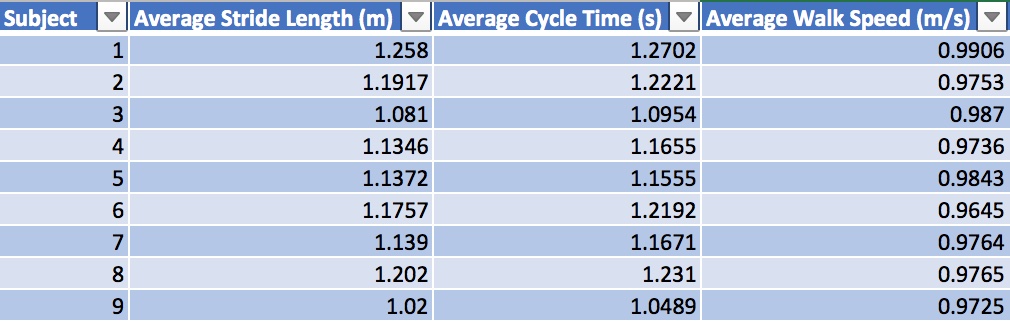, width=4.5in}
	\caption{Calculated stride length values using our method.  Included is the average walk speed (stride length divided by cycle time) which we use to test the consistency and accuracy of different methods.  The treadmill was set to run at 1 m/s and has a margin of error of about 1 percent.  The mean of our calculated walk speeds is 0.9779 m/s, which is a 2.2 percent error.}
	\label{stridelengthtable}
\end{figure*}

The data we used are of 9 able-bodied subjects walking for 60 seconds on a treadmill set at 1 m/s with no incline.  The data was captured with Vicon sensors and cameras at the University of Texas Southwestern and was filtered before we received it at our lab at the University of Texas at Dallas.  We used MATLAB (The MathWorks, Inc., Natick, Massachusetts, United States) to calculate stride lengths and cycle times (cycle time is the duration of a stride: we calculated this as the time between consecutive same-foot heel strikes) for every stride in each data set, and the average values of the parameters are presented in Figure \ref{stridelengthtable}.

Additionally, we calculated walking speed for each stride.  This was calculated as (stride length)/(cycle time) \cite{Whittle}, and is included in Figure \ref{stridelengthtable}.  We use this parameter to compare the accuracy and consistency of different stride length methods.  The reason we use this parameter for comparison is that it is the variable that most resembles a control; we did not have a chance to manually measure each stride length to get an accurate value, so we chose to compare our results to the most consistent variable available.  It is worth noting that walking speed does have potential error: this particular treadmill has a tolerance of about 1 percent, and others have discussed that walking speed can vary from the set treadmill speed \cite{Souza16}.  However, since the potential error is minimal, walking speed is a good variable to use for checking whether our stride length calculation algorithm is close and realistically reliable.

Our calculated walk speeds had a mean of 0.9779 m/s and a standard deviation of 0.0081 m/s.  The low standard deviation suggests that this method works consistently for different subjects, and the error of 2.2\% is acceptable in regards to gait analysis \cite{qi2014estimation}, suggesting that this method is sufficiently accurate.  

\section{Discussions}
\label{sec-discussions}

%Discuss results
There is a possible threat to validation that is worth discussing. Due to the mechanics of walking and the methodology of our step length calculation, there exists a potential for errors; depending on the subject, at the moment of heel-strike the heel of the hind foot may be lifted at a large enough angle to cause a small offset in the horizontal position of the heel. There also exists a limitation in our experiment; since we only tested our method with data from able-bodied subjects walking on 0 incline at 1 m/s, we cannot be sure that the same method of calculation will provide a similar accuracy for all situations of human locomotion. 

Additionally, we discovered a potential misinterpretation of heel marker data that could lead to faulty stride length calculations. We did not find many papers that made this exact mistake (as aforementioned, Qi et al. \cite{qi2014estimation} used a similar calculation to the one discussed below), but we think it is worth addressing since it seems intuitive and might be an easy mistake to make.

\subsection{Possible errors and limitations}
\label{subsec-limitations}

First, the calculated stride lengths are all less than the expected value of 1 m/s.  The percent of error is not large, so it is possible that this observation is insignificant and caused by minor errors in data or treadmill performance.  However, there may be a legitimate theoretical explanation for the calculated values being less than the expected value.  This error most likely occurs during the step length calculation: at the moment of heel strike, the heel of the opposite foot is usually slightly off the ground.  The horizontal distance lost in this lift is not much compared to the overall step length and might be generally negligible, but if a subject tends to lift their opposite heel high during heel strike, then there may be a not-insignificant shift in horizontal position that could lead to a decrease in measured step length.  Further research should be done to determine whether this error is a legitimate problem that potentially interferes with the practical applications of stride length.  Additionally, the proposed method has only been tested for 0 incline walking, so other situations such as running or inclined walking may yield different results.

Furthermore, the number of subjects in this study is low, and we did not have a control group that used an established accurate method of stride length calculation such as the video-analysis method discussed in \cite{Murray87}.  The reason for the lack of ideal experimental data is that this data was not originally collected to test different methods of stride length calculations; our lab builds and studies robotic legs, so gait calculation is an intermediate step as opposed to the main objective for data collection.  However, while working on gait calculations we deduced this new approach that is more accurate and may be worth sharing, so we used the data available to demonstrate the merits of this method.

\subsection{Total distance heel travels does not equal stride length}
\label{subsec-mistake}

The most intuitive approach is a faulty one: stride length is defined as the distance a foot travels from one heel strike to the next, so intuitively it seems that the total distance the heel travels during a stride is the stride length.  This is accurate for grounded walking; however, this is not the case for treadmill walking.  As others have mentioned, methods that work accurately for grounded walking may not necessarily translate to treadmill walking \cite{DEWITT20103067}.  

\begin{figure}[b]
	\centering
	\epsfig{file=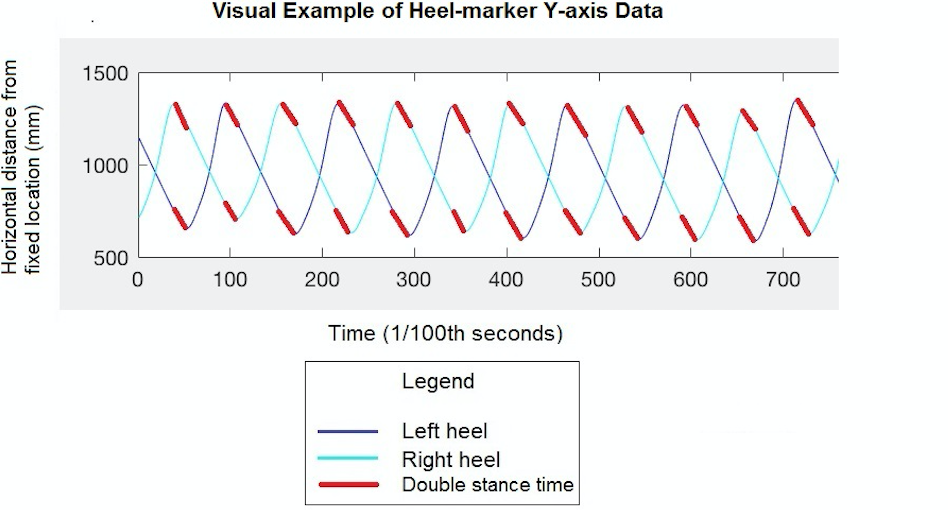, width=3.3in}
	\caption{Here, double stance time (DST) is highlighted.  During DST, both heels are moving backwards since the feet are planted while the treadmill keeps moving. This backwards drift is recorded in the data, but it does not contribute any forward progress since both feet are stationary relative to the platform.}
	\label{doublestancetime}
\end{figure}

\begin{figure*}
	\centering
	\epsfig{file=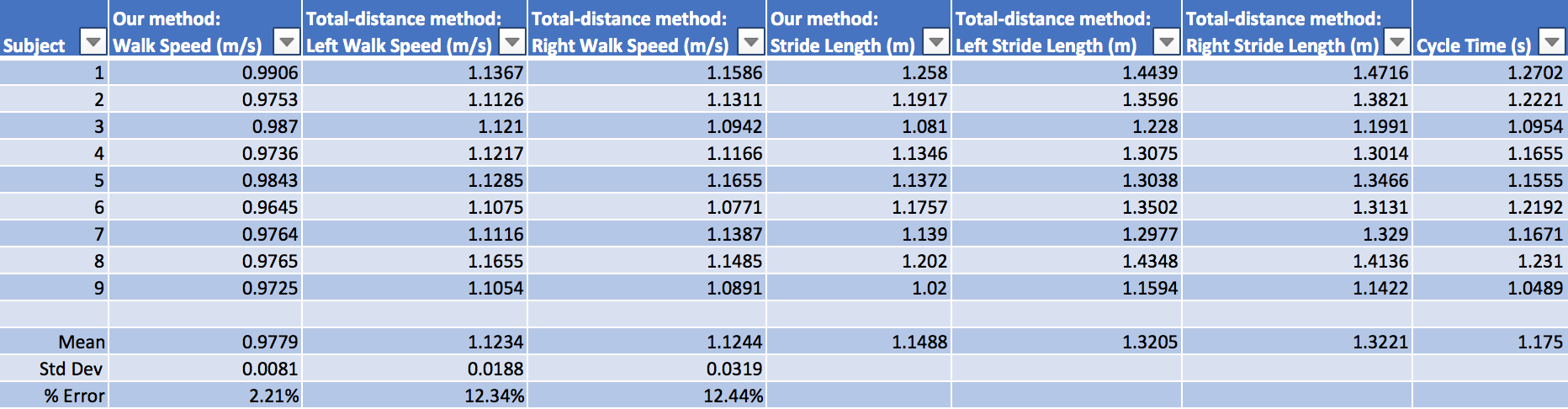, width=6in}
	\caption{The faulty total-distance-traveled method of stride length calculation resulted in average errors of over 12\% (compared to the expected walking speed of 1 m/s) and had standard deviations that were larger than that of our method.}
	\label{comparemethods}
\end{figure*}

To elaborate, the heel marker data collected from treadmill walking can potentially be misinterpreted.  When the heel-marker position is increasing, this is a swing, since the foot is moving forward.  When the horizontal position is decreasing, this is a stance, since the treadmill is bringing the planted foot backwards.  The forward swing is legitimate distance that the foot covers.  However, the backwards movement is not indicative of anything for that specific foot.  Regardless of how the moving platform affects the data, the planted foot is not walking any distance while in stance.  One may argue that while one foot is in stance time on the treadmill, the other foot is in swing, so the displacement of the stationary foot during that time should average out to equal the distance covered by the other foot's swing.  There are two problems with this idea: one is that swing distance is not equivalent to step length, so summing two swing distances is not equal to a stride length.  Another problem is that a fraction of a foot's stance time is actually double stance time, when neither foot is making any forward progress. This is the main issue.  The reason the backwards displacement of the heel does not indicate anything important is that during double stance time, the positional heel data shows distance traveled but the foot is not actually doing anything (see Figure \ref{doublestancetime}).  Thus if we use total distance traveled as our measurement, we are including the heel's backwards displacement during double stance time which does not contribute any distance to the stride length.  

One may argue that the distance lost during double stance time must be made up eventually, otherwise the person would keep drifting backwards off the treadmill.  Although it is true that the foot makes up the lost distance, we only need to measure that distance during the forward swing; we do not need to measure it twice by including the backwards drift of double stance time.  Consider the following metaphor: if we measure distance driven by a car, we measure the distance the car moves forward.  Suppose the car drives 10 meters, then a person manually pushes the car back 5 meters, and the car drives 5 more meters: we conclude that the car drove 15 meters.  The distance we measure is the forward motion of the car: the car was not driving any distance when the person pushed the car back, so we cannot say the car drove 20 meters of distance.  The car does make up the 5 meters it lost; this made up distance contributes to the total distance driven.  However, we do not need to include this 5 meters twice.  Similarly, during double stance time the treadmill pushes the person backwards, and the person has to make up this distance.  However, we only need to measure this distance once, meaning that the heel data measured during double stance time is extraneous when considering stride length.  Therefore a total-distance-traveled calculation of the heel marker data would theoretically yield a larger stride length than in actuality.  

To illustrate its noticeable error, we calculated stride length and walking speed using the total-distance method.  To calculate total-distance traveled by the heel between heel strikes, we subtracted the local minimum heel position from the previous local maximum, then subtracted the local minimum from the next local maximum, then summed the two values.  We repeated this process for each stride and averaged the values, and we also calculated walking speed for comparison.  For some subjects, the results obtained from the left and right data were slightly different, so in Figure \ref{comparemethods} we included separate calculations for the two sides.  As predicted, the calculated walking speeds from the total-distance method are much larger than the expected value (treadmill speed) of 1 m/s; furthermore, the percent error and inconsistency (evidenced by the standard deviation of the walking speeds) of the total-distance method are noticeably worse than those of our approach.  The average error of over 12\% is also much too large to be explainable by the treadmill's aforementioned tolerance.  These all suggest that the total-distance method is not a viable method for computing stride length for treadmill walking.

\section{Conclusion}
\label{sec-conclusion}

In this paper we presented a new approach to calculating stride length from positional heel-marker data for treadmill walking.  Stride length is geometrically equivalent to the sum of two consecutive step lengths, and step length can be directly calculated from heel-marker data.  Thus we computed stride length in this manner: finding direct calculations of step lengths at the times of heel strike, then summing consecutive step lengths together. Overall, our method of calculating stride length is more accurate than other digital methods we found in literature. Our digital method is also an improvement in practicality over previously established video-based methods because digital methods do not require manual calculations for each stride. Additionally, this paper provides an in-depth discussion and analysis of the stride length calculation process, since such an explanation for stride length can be difficult to find in other literature.

Additionally, the ability to accurately calculate stride length from horizontal heel data further increases the versatility of this particular type of kinematic data.  Horizontal heel data can already lead to the detection of heel strike and toe-off events \cite{DESAILLY200976}, which are very important for finding useful gait parameters such as stance and swing times \cite{BRUENING2014472}.  The ability to use kinematic data such as horizontal heel displacement to determine so many spatial-temporal gait parameters can be useful, especially if sufficient kinetic data is difficult to obtain \cite{Hendershot20164146}.  Thus the method of stride length calculation presented in this study adds a new tool to the already important toolbox of heel marker data, allowing for more practical applications in gait analysis.

However, our method still has potential for improvement.  Further research can be done to investigate the limitations of this study. A study can be conducted to apply the method of calculation to multiple types of gait data, such as abnormal gait, inclined walking, or running.  A study can also be conducted to test the accuracy of the method compared to an accurate and established method such as the video-analysis stride length calculation; such a study would allow for a clearer conclusion of whether the potential horizontal offset at heel strike affects the accuracy of the step length approach.

\section{Acknowledgements}
\label{sec-acknowledgements}

The author would like to thank Robert Gregg, Ph.D. and his Locomotor Control Systems Laboratory at the University of Texas at Dallas for giving him an opportunity to work with the lab as an undergraduate research intern.  The author would especially like to thank David Quintero for taking him in as a personal intern and mentoring him through the process of research.

%\section*{Acknowledgment}
%more thanks here



%\bibliographystyle{IEEEtran}

\scriptsize

\bibliographystyle{IEEEtran}
\bibliography{kevin_supakkul}

\end{document}